\documentclass[a4paper,11pt]{article}
\usepackage{latexsym}
\usepackage{amssymb}
\usepackage{amsfonts}
\usepackage{amsmath}
\usepackage{indentfirst}
\usepackage{graphicx}
\usepackage{epsfig}
\usepackage{amsthm,xspace}
\usepackage{graphics,pst-tree}
\usepackage{array,multirow}
\usepackage{eufrak}
\newtheorem{prop}{Proposition}
\newtheorem{teo}{Theorem}

\newcommand{\pat}{\mathfrak{p} }
\newcommand{\fine}{$\hfill \square$}
\newcommand{\barx}{\bar{x}}
\author{Stefano Bilotta \and Elisabetta Grazzini
\and Elisa Pergola \thanks{Communicating author: \emph{Phone:
+390554237458\ Fax: +390554237436\ e-mail:
elisa@dsi.unifi.it}.\newline \indent E-mails:
\texttt{\{bilotta,elisa,pinzani@dsi.unifi.it\},
elisabetta.grazzini@unifi.it }} \and Renzo Pinzani}

\title{Generation of binary words avoiding alternating patterns}
\date{}
\begin{document}
\maketitle
\begin{center}
\vspace{-1cm} \noindent \small{Dipartimento di Sistemi e
Informatica, Universit\`a di Firenze\\ Viale G. B. Morgagni 65,
50134 Firenze, Italy.}
\end{center}

\begin{abstract}In this paper we propose an algorithm to
generate binary words with no more 0's than 1's having a fixed
number of 1's and avoiding the pattern $(10)^j1$ for any fixed $j
\geq 1$. We will prove that this generation is exhaustive, that
is, all such binary words are generated.
\end{abstract}
\textbf{Keywords:} Binary words, Pattern avoiding, Exhaustive
generation.
\section{Introduction}
The problem of determining the appearance of a fixed
\emph{pattern} in long sequences of observation is relevant in
many scientific problems.

For example, in the area of computer network security, the
detection of intrusions, which become increasingly frequent, is
very important. Intrusion detection is primarily concerned with
the detection of illegal activities and acquisitions of privileges
that cannot be detected by information flow and access control
models. There are several approaches to intrusion detection, but
recently this subject has been studied in relation to pattern
matching (see \cite{AA02,FSV06,KS94}).

This leads to the study of the construction of particular words
avoiding a given pattern in an alphabet $\Sigma$. The present
paper aims to be a contribution in this direction.

Let $F \subset \{0,1\}^*$ be the set of binary  words $\omega$
such that $|\omega|_0 \leq |\omega|_1$, for any $\omega \in F$,
$|\omega|_0$ and $|\omega|_1$ corresponding to the number of 0's
and 1's in the word $\omega$, respectively. In this paper we study
the construction of the subset $F^{[\pat]} \subset F$ of binary
words excluding a given pattern $\pat = p_0 \ldots p_{\ell-1} \in
\{0,1\}^\ell$, that is a word $\omega \in F^{[\pat]}$ if and only
if it does not contain a sequence of consecutive indices $i, i+1,
\ldots,i+\ell-1$ such that $\omega_i,\omega_{i+1} \ldots
\omega_{i+\ell-1} = p_0 p_1\ldots p_{\ell-1}$.

If we consider the set of binary words without any restriction,
the defined language is regular and we can refer to using
classical results (see, e.g., \cite{GO80,GO81,SF95}). When the
restriction to words with no more 0's than 1's is valid, the
language $F^{[\mathfrak{p}]}$ is not a regular one and it becomes
more difficult to deal with. For example, in order to generate the
language $F^{[\mathfrak{p}]}$ for each forbidden pattern $\pat$ an
``ad hoc'' grammar should be defined. Our aim is to determine a
constructive algorithm suggesting a more unified approach which
makes it possible to generate all binary words in the class
$F^{[\mathfrak{p}]}$.

In this paper we show how to obtain all binary words belonging to
$F$ and avoiding the pattern $\pat = (10)^j1$, for any fixed $j
\geq 1$.

We \cite{BMPP11} introduced an algorithm for the construction of
all binary words in $F$ having a fixed number of 1's and excluding
those containing the forbidden pattern $1^{j+1}0^j$, for any fixed
$j \geq 1$. That algorithm generates all the words in $F$ then
eliminates those containing the forbidden pattern. Basically, the
construction marks in an appropriate way the forbidden patterns in
the words and generates $2^C$ copies of each word having $C$
forbidden patterns such that the $2^{C-1}$ instances containing an
odd number of marked forbidden pattern are annihilated by the
other $2^{C-1}$ instances containing an even number of marked
forbidden patterns. For example, the words $00110\overline{110}$
and $00\overline{110}110$, containing two copies of the forbidden
pattern $\pat=110$, (the marked forbidden patterns are over-lined)
are eliminated by the words $00110110$ and
$00\overline{110}\overline{110}$, respectively.

This is possible since no prefix of $\pat=1^{j+1}0^j$ is also a
suffix of $\pat$, that is the forbidden patterns do not overlap
and so they are univocally identified inside the words.

Then, the algorithm in \cite{BMPP11} cannot be used to generate
the words in $F^{[\mathfrak{p}]}$ when $\pat = (10)^j1$ since the
forbidden patterns may overlap inside the words. For example, in
$\omega = 110101010$ there are two overlapping copies of the
forbidden pattern $\pat = (10)^21$. So, we propose a new algorithm
that generates right the words in $F$ avoiding the forbidden
pattern $\pat = (10)^j1$, for any fixed $j \geq 1$.

The paper is organized as follows. In Section~\ref{sec:basic} we
give some basic definitions and notation. In particular, we recall
how every binary word can be represented as a path on the
Cartesian plane.

In Section~\ref{sec:algo} we give a construction, according to the
number of 1's, for the set of binary words excluding the pattern
$\pat = (10)^j1$, for any fixed $j \geq 1$, and such that the
number of 0's in each word is inferior to or the same as the
number of 1's.

In Section~\ref{sec:proof} we prove that the construction given in
Section~\ref{sec:algo} allows us to obtain an exhaustive and
univocal generation of such binary words having $n$ 1's.

\section{Basic definitions and notation}\label{sec:basic} Let $F \subset \{0,1\}^*$ be the set of
binary  words $\omega$ such that $|\omega|_0 \leq |\omega|_1$, for
any $\omega \in F$, $|\omega|_0$ and $|\omega|_1$ corresponding to
the number of 0's and 1's in the word $\omega$, respectively. In
this paper we study the construction of the subset $F^{[\pat]}
\subset F$ of binary words excluding a given pattern $\pat =
(10)^j1$, for any fixed $j \geq 1$.

Given $|\omega| = |\omega|_0 + |\omega|_1$ the \emph{length} of
$\omega \in F$, we denote by $\omega^h$, ($h
>0$), the word with length $h \cdot |\omega|$ obtained by linking $\omega$ to itself $h$
times, that is
$\omega^h = \underbrace{\omega\,\omega \cdots \omega}_{h}$
and
 $\omega^0 = \varepsilon$, $\varepsilon$ being the empty word.

\begin{figure}[!htb]
\begin{center}
\epsfig{file=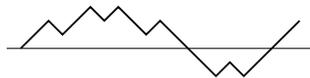, width =1.6in, clip=}
\caption{\small{The path representing  $\omega
=11011010010000101111$}} \label{fig:path}\vspace{-15pt}
\end{center}
\end{figure}

Each word $\omega \in F$ can be naturally represented as a path on
the Cartesian plane by associating a \emph{rise} (or \emph{up})
\emph{step}, defined by (1,1) and indicated by $x$, with each bit
1 in $\omega$ and a \emph{fall} (or \emph{down}) \emph{step},
defined by (1,-1) and indicated by $\barx$, with each bit 0 in
$\omega$. For example, the word $\omega =11011010010000101111$ is
represented by the path $\gamma = xx\barx xx\barx x\barx\barx
x\barx\barx\barx\barx x\barx xxxx$ (see Figure~\ref{fig:path}). An
\emph{up-down} step is the sequence $x\barx$.

From now on, we refer interchangeably to words or their graphical
representation on the Cartesian plane, that is paths. So by
$F^{[\pat]}$ we denote both the set of pattern $\pat$ avoiding
binary words and the set of corresponding paths.

\noindent In the rest of this paper, a path is defined as:
\begin{itemize}
\item[-] \emph{primitive} if it begins and ends at ordinate 0 and
remains strictly above the $x$-axis,

\item[-] \emph{positive} if it begins
 at ordinate 0 and remains above or on the $x$-axis,
\item[-] \emph{negative} if it begins and ends at ordinate 0 and
remains below or on the $x$-axis (remark that a negative path in
$F$ necessarily ends at ordinate 0),

\item[-] \emph{strongly negative} if it begins and ends at
ordinate -1 and remains below or on the line $y=-1$,

\item[-]  \emph{underground} if it ends with a negative suffix.
\end{itemize}

The \emph{complement} of a path $\varphi$ is
 the path $\varphi^c$ obtained from $\varphi$ by switching rise and fall
 steps.

\section{A construction for the set
$F^{[\pat]}$}\label{sec:algo}

In this section we show the constructive algorithm to generate the
set $F^{[\pat]}$, $\pat = (x\barx)^jx= (10)^j1$ for any fixed $j
\geq 1$, according to the number of rise steps, or equivalently to
the number of 1's. Given a path $\omega \in F^{[\pat]}$ with $n$
rise steps, we generate a given number of paths in $F^{[\pat]}$
with $n+h$ rise steps, $1 \leq h \leq j$, by means of constructive
rules. The number and the shape of the generated paths depend on
the ordinate $k$ of the endpoint of $\omega$ and on its suffix.
With regard to $k$, we can point out three cases: $k=0$, $k=1$ and
$k \geq 2$, while as for the suffix we consider whether it is
equal to $(x\barx)^j$ or not. When $k=0$, we must pay attention
also to the case in which $\omega$ is an underground path ending
with the pattern $(x\barx)^{j-1}x$.

As we will show further on, for each $\omega \in F^{[\pat]}$ such
that $k=0$ or $k \geq 2$, the generating algorithm produces two or
more positive paths and one underground path with $n+h$ rise
steps, $1 \leq h \leq j$, while, when $k=1$, it produces only one
positive path with $n+h$ rise steps.

\noindent Let us denote by $\omega_{|k}$ a path with endpoint at
ordinate $k$.

The generating algorithm of the class  $F^{[\pat]}$ with $\pat =
(x\barx)^jx = (10)^j1$, for any fixed $j \geq 1$, is described in
the following sections. The constructive rules related to the
special cases in which the suffix of $\omega$ is $(x\barx)^j$ or
$(x\barx)^{j-1}x$ are described in Sections
\ref{sec:positivesuffix} and \ref{sec:negativesuffix},
respectively, while in Section \ref{sec:simplecase} we examine all
the other simple cases.

\noindent The starting point of the algorithm is the empty word
$\varepsilon$.
\subsection{Simple cases}\label{sec:simplecase} In this section we
describe the constructive rules to be applied when the suffix of
$\omega$ is neither $(x\barx)^j$ nor $(x\barx)^{j-1}x$. We point
out three cases for the ordinate $k$ of the endpoint of $\omega$:
$k=0$, $k=1$ and $k \geq 2$.
\begin{description}

\item [$k=0$.] A path $\omega \in F^{[\pat]}$, with $n$ rise steps
and such that its endpoint has ordinate 0, generates, for any $h$,
$1 \leq h \leq j$, three paths with $n+h$ rise steps: a path
ending at ordinate 1 by adding to $\omega$ a rise step and a
sequence of $h-1$ up-down steps; a path ending at ordinate 0 by
adding to $\omega$ a rise step, a sequence of $h-1$ up-down steps
and a fall step, and an underground path obtained by the one
generated in the previous step and mirroring on $x$-axis its
rightmost primitive suffix.

Figure~\ref{fig:zero1} shows the above described operations; the
number above the right arrow corresponds to the value of $h$. Both
in this figure and in the following ones we consider $j=4$, that
is $\pat=(x\barx)^4x = (10)^41$.
\begin{figure}[!htb]
\begin{center}
\epsfig{file=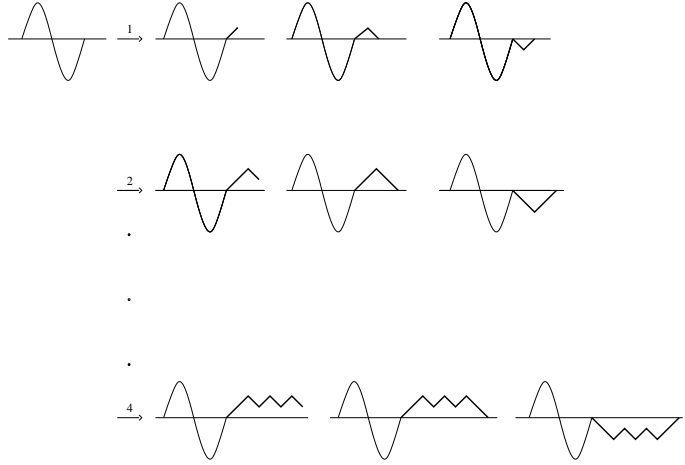, width =3.5in, clip=} \caption{\small{The
paths generated by $\omega_{|0}$}} \label{fig:zero1}\vspace{-15pt}
\end{center}
\end{figure}

\noindent Therefore
\begin{equation}\label{alfa}
\omega_{|0} \Rightarrow \left\{
\begin{array}{l}
\omega_{|0}x\,(x\barx)^{h-1}\\
\omega_{|0}x(x\barx)^{h-1}\barx\\
\omega_{|0}\barx (\barx x)^{h-1}x\\
\end{array}
\right.
\end{equation}

\item [$k=1$.] A path $\omega \in F^{[\pat]}$, with $n$ rise steps
and such that its endpoint has ordinate 1, generates, for any $h$,
a path with $n+h$ rise steps with endpoint at ordinate 2 obtained
by adding to $\omega$ a rise step and a sequence of $h-1$ up-down
steps (see Figure~\ref{fig:uno1}).

\begin{figure}[!htb]
\begin{center}
\epsfig{file=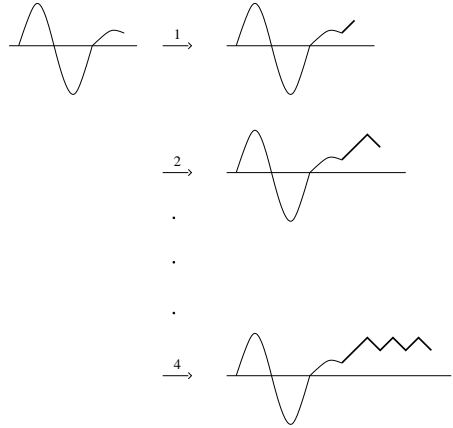, width =2.3in, clip=} \caption{\small{The
paths generated by $\omega_{|1}$}} \label{fig:uno1}\vspace{-15pt}
\end{center}
\end{figure}

\noindent Therefore
\begin{equation}\label{beta}
\omega_{|1}\Rightarrow \omega_{|1}x(x\barx)^{h-1}
\end{equation}

\item [$k \geq 2$.] A path $\omega \in F^{[\pat]}$, with $n$ rise
steps and such that its endpoint has ordinate $k$, $k \geq 2$,
generates, for any $h$, $k+2$ paths with $n+h$ rise steps: a path
ending at ordinate $(k+1)$ by adding to $\omega$ a rise step and a
sequence of $h-1$ up-down steps; $k-1$ paths ending at ordinate
$(k-1), (k-2), \ldots, (1)$, respectively,  by adding to $\omega$
a rise step, a sequence of $m$, $2 \leq m \leq k$, fall steps and
a sequence of $h-1$ up-down steps; a path ending at ordinate 0 by
adding to $\omega$ a rise step, a sequence of $k$ fall steps, a
sequence of $h-1$ up-down steps and a fall step, and an
underground path which will be described in
Section~\ref{sec:under}. Figure~\ref{fig:due1} shows the above
described operations.

\begin{figure}[!htb]
\begin{center}
\epsfig{file=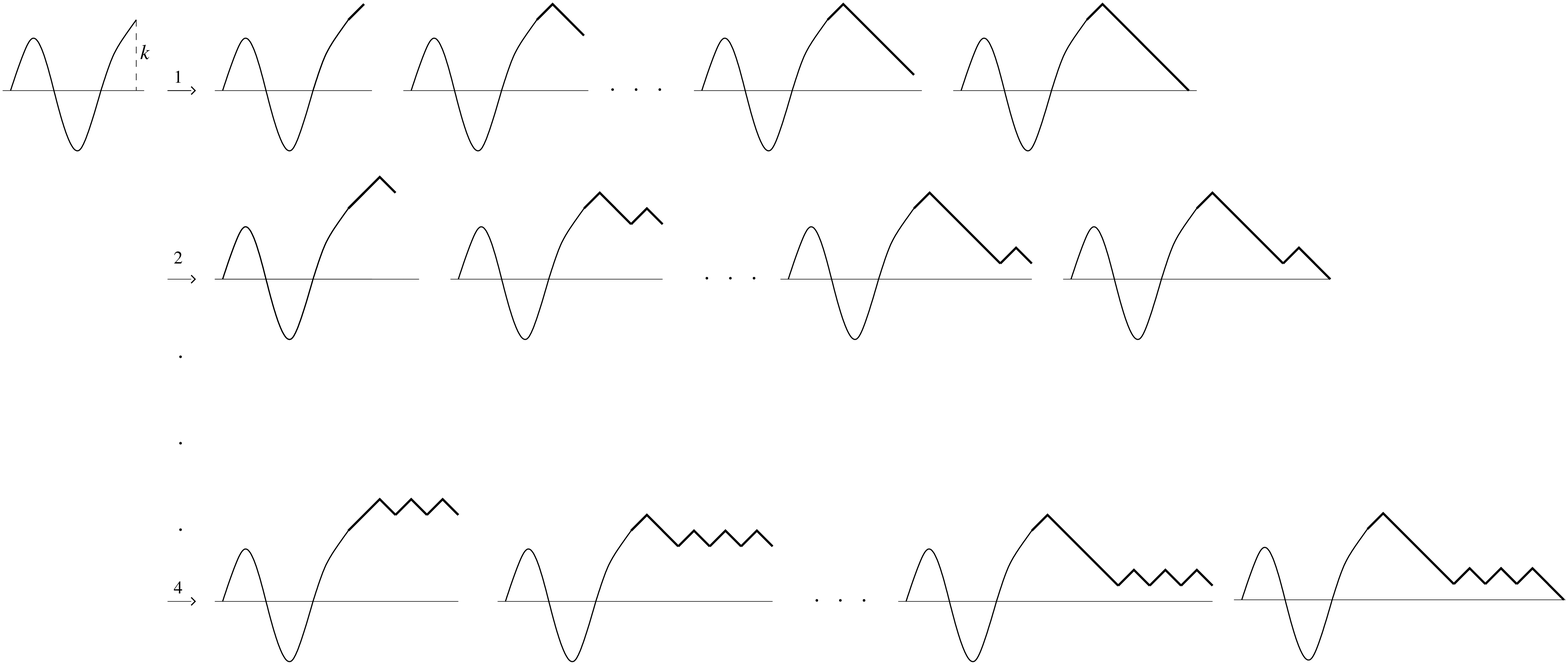, width =5.3in, clip=} \caption{\small{The
paths generated by $\omega_{|k}, k\geq 2$}}
\label{fig:due1}\vspace{-15pt}
\end{center}
\end{figure}

Therefore

\begin{equation}\label{gamma}
\omega_{|k} \Rightarrow \left\{
\begin{array}{ll}
\omega_{|k}x(x\barx)^{h-1}\\
\omega_{|k}x(\barx)^m(x\barx)^{h-1}&2 \leq m \leq k\\
\omega_{|k}x(\barx)^k(x\barx)^{h-1}\barx\\
\end{array}
\right.
\end{equation}
\end{description}

At this point it is clear that:
\begin{enumerate}
\item when the path $\omega$ ends with the suffix $(x\barx)^j$ the
paths obtained by means of the constructions (\ref{alfa}),
(\ref{beta}) and (\ref{gamma}) contain the forbidden pattern
$\pat=(x\barx)^jx$. So, we will act as described in
Section~\ref{sec:positivesuffix};

\item when $\omega$ is an underground path ending with the pattern
$(x\barx)^{j-1}x$, some  paths generated by means of the above
constructions might contain the forbidden pattern
$\pat=(x\barx)^jx$. So, we will follow a different procedure
described in Section~\ref{sec:negativesuffix}.

\end{enumerate}
\subsection{Paths ending with
$(x\barx)^j$}\label{sec:positivesuffix} Even when the path
$\omega$ ends with the suffix  $(x\barx)^j$, the number and the
shape of the generated paths depend on the ordinate $k$ of the
endpoint of $\omega$.  Let $\varrho=(x\barx)^j$ be the suffix of
$\omega$.
\begin{description}

\item [$k=0$.] A path $\omega \in F^{[\pat]}$, with $n$ rise steps
and such that its endpoint has ordinate 0, generates, for any $h$,
$1\leq h \leq j$, three paths with $n+h$ rise steps (see
Figure~\ref{fig:zero2}): a path ending at ordinate 1, by inserting
a sequence of $h-1$ up-down steps and a rise step on the left of
$\varrho$; a path ending at ordinate 0, by inserting a sequence of
$h-1$ up-down steps and a rise step on the left of $\varrho$ and
adding a fall step at the end of $\omega$, and an underground
path, obtained by mirroring on $x$-axis the rightmost primitive
suffix of the path generated at the previous step. Therefore

\begin{equation}\label{alfa1}
\omega_{|0}\varrho \Rightarrow \left\{
\begin{array}{l}
\omega_{|0}(x\barx)^{h-1}x\varrho\\
\omega_{|0}(x\barx)^{h-1}x\varrho\,\barx\\
\omega_{|0}(x\barx)^{h-1}\barx \barx (x\barx)^{j-1}xx\\
\end{array}
\right.
\end{equation}

\begin{figure}[!htb]
\begin{center}
\epsfig{file=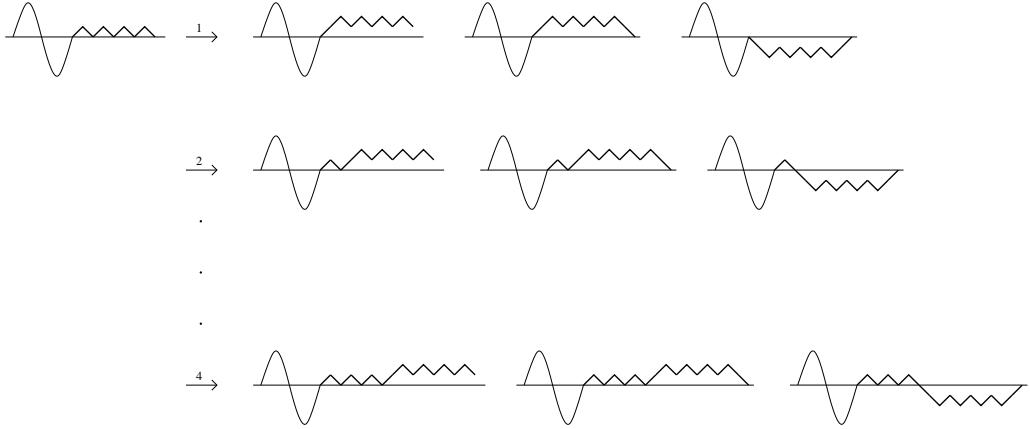, width =5.3in, clip=} \caption{\small{The
paths generated by $\omega_{|0}(x\barx)^j$}}
\label{fig:zero2}\vspace{-15pt}
\end{center}
\end{figure}

\item [$k=1$.] A path $\omega \in F^{[\pat]}$, with $n$ rise steps
and such that its endpoint has ordinate 1, generates, for any $h$,
a path with $n+h$ rise steps with endpoint at ordinate 2, obtained
by inserting a sequence of $h-1$ up-down steps and a rise step on
the left of the suffix $\varrho$ (see Figure~\ref{fig:uno2}).
Therefore

\begin{equation}\label{beta1}
\omega_{|1}\varrho \Rightarrow \omega_{|1}(x\barx)^{h-1}x\,\varrho
\end{equation}

\begin{figure}[!htb]
\begin{center}
\epsfig{file=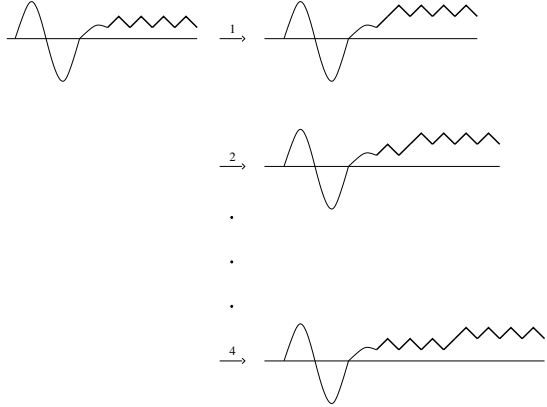, width =2.8in, clip=} \caption{\small{The
paths generated by $\omega_{|1}(x\barx)^j$}}
\label{fig:uno2}\vspace{-15pt}
\end{center}
\end{figure}

\item [$k \geq 2$.] A path $\omega \in F^{[\pat]}$, with $n$ rise
steps and such that its endpoint has ordinate $k$, $k \geq 2$,
generates, for any $h$, $k+2$ paths with $n+h$ rise steps (see
Figure~\ref{fig:due2}): a path  ending at ordinate $(k+1)$, by
inserting a sequence of $h-1$ up-down steps and a rise step on the
left of the suffix $\varrho$; $k-1$ paths ending at ordinate
$(k-1), (k-2), \ldots, (1)$, respectively, by inserting a sequence
of $h-1$ up-down steps, a rise step and a sequence of $m$, $2 \leq
m \leq k$, fall steps on the left of $\varrho$; a path ending at
ordinate 0, by inserting a sequence of $h-1$ up-down steps, a rise
step and a sequence of $k$ fall steps on the left of $\varrho$,
and then adding a fall step at the end of $\omega$, and an
underground path which will be described in
Section~\ref{sec:under}. Therefore

\begin{equation}\label{gamma1}
\omega_{|k}\varrho \Rightarrow \left\{
\begin{array}{ll}
\omega_{|k}(x\barx)^{h-1}x\varrho\\
\omega_{|k}(x\barx)^{h-1}x(\barx)^m\,\varrho \quad 2 \leq m \leq k\\
\omega_{|k}(x\barx)^{h-1}x\,(\barx)^k\,\varrho\,\barx\\
\end{array}
\right.
\end{equation}

\begin{figure}[!htb]
\begin{center}
\epsfig{file=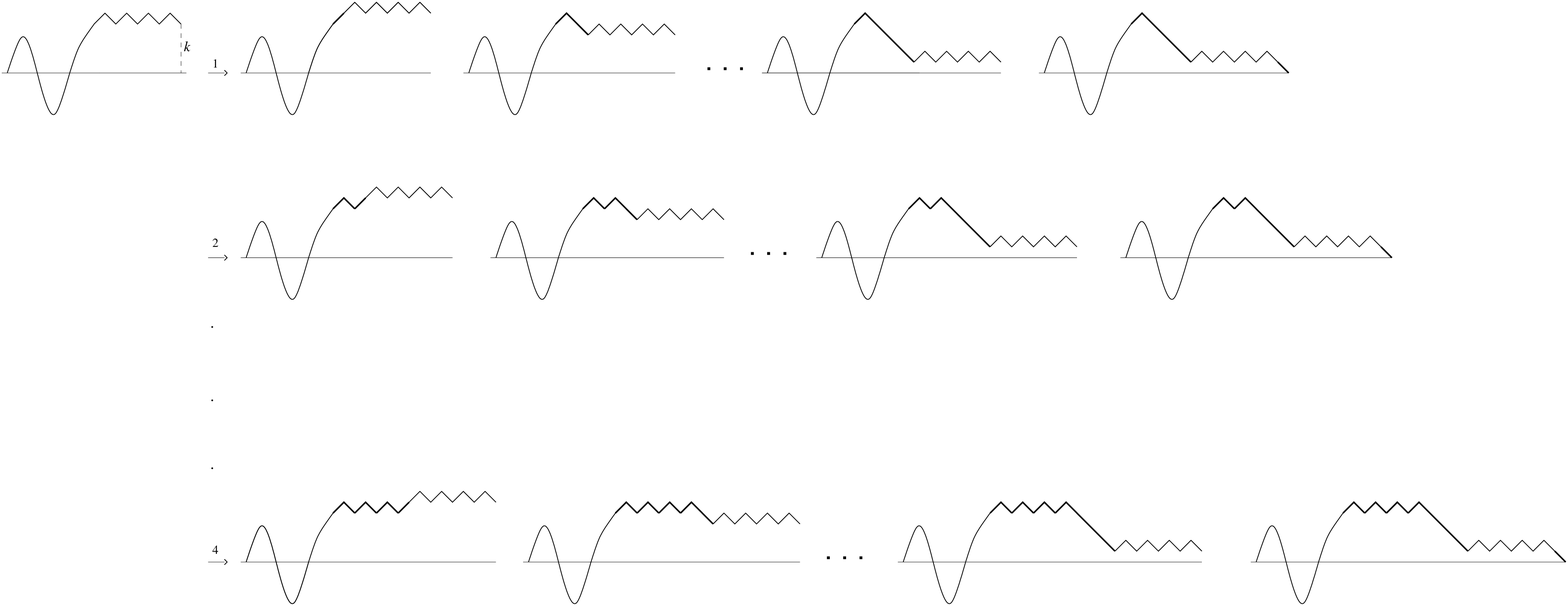, width =6.3in, clip=}\caption{\small{The
paths generated by $\omega_{|k}(x\barx)^j, k\geq 2$}}
\label{fig:due2}\vspace{-15pt}
\end{center}
\end{figure}

\end{description}

\subsection{Paths ending with
$(x\barx)^{j-1}x$}\label{sec:negativesuffix} The paths $\omega \in
F^{[\pat]}$ ending on the $x$-axis with the sequence
$(x\barx)^{j-1}x$ have the following shape
$$\omega_{|0} =\mu\,\barx \,\eta\,x\,(\barx x)^{j-1}$$
where $\mu$ is a path ending on the $x$-axis and $\eta$ is either
the empty path $\varepsilon$ or is a strongly negative path.

The constructions applied to paths ending at ordinate 0 described
in (\ref{alfa}) (see Figure~\ref{fig:zero1}) can be used even for
the paths ending with the sequence $(x\barx)^{j-1}x$, when $h \geq
2$, or to generate the paths ending at ordinate 1 or on the
$x$-axis with a positive suffix, when $h=1$. Nevertheless, when
$h=1$, by applying the construction, we obtain an underground path
which contains the forbidden pattern $\pat= (x\barx)^jx$.

Therefore if the path ends with the sequence $(x\barx)^{j-1}x$ and
$h=1$, in order to generate the underground path we proceed as
follows. Two cases must be taken into consideration.

\begin{description}

\item[1) $\mu$ does not end with a peak $x\barx$.] The underground
path generated from $\omega_{|0} =\mu\,\barx \,\eta\,x\,(\barx
x)^{j-1}$ is obtained by adding the path $\barx x$ to
$\omega_{|0}$, mirroring on $x$-axis the rightmost suffix $(\barx
x)^j$ of $\omega_{|0}\barx x$ and shifting the sequence
$(x\barx)^j$ between $\mu$ and the sub-path $\barx \,\eta \,x$.

So the path $\omega_{|0} =\mu\,\barx \,\eta\,x\,(\barx x)^{j-1}$
generates the underground path $\mu\,(x\barx)^j\,\barx \,\eta \,x$
(see Figure~\ref{fig:casoa}). It should be noticed that this
construction applies to $\omega$ even if $\mu = \varepsilon$.

\begin{figure}[!htb]
\begin{center}
\epsfig{file=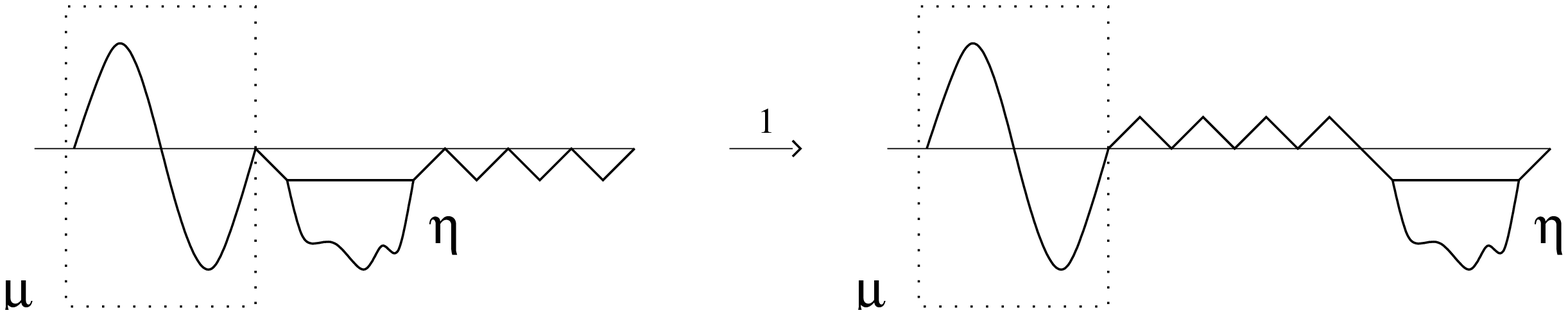, width =3.3in,
clip=}\caption{\small{The underground path generated by
$\omega_{|0}$ in the case 1)}}\label{fig:casoa}\vspace{-15pt}
\end{center}
\end{figure}

\item[2) $\mu$ ends with a peak $x\barx$.] When the path $\mu$
ends with a peak $x\barx$, that is $\mu= \mu'x\barx$, the
insertion of the sequence $(x\barx)^j$ between $\mu$ and the
sequence $\barx \, \eta \,x$ produces the forbidden pattern $\pat
= (x\barx)^jx$. Let us consider the following subcases: $\eta \neq
\varepsilon$ and $\eta = \varepsilon$.
\begin{description}
\item [2.1) $\eta  \neq \varepsilon$.] The underground path is
obtained by performing on $\omega_{|0} =\mu' x \barx \barx
\eta\,x\,(\barx x)^{j-1}$ the following operations: shifting the
rightmost peak $x\barx$ of $\mu$ to the right of the sub-path
$\barx \,\eta \,x$, mirroring on $x$-axis the sequence $(\barx
x)^{j-1}$ and adding to such path the steps $\barx\, x$.

So, when $h=1$, the underground path with negative suffix
generated by $\omega_{|0} =\mu'\,x\,\barx \, \barx \,\eta
\,x\,(\barx x)^{j-1}$ is $\mu'\, \barx \,\eta \,x\,(x\barx)^{j}\,
\barx \, x$ (see Figure~\ref{fig:casob1}).

\begin{figure}[!htb]
\begin{center}
\epsfig{file=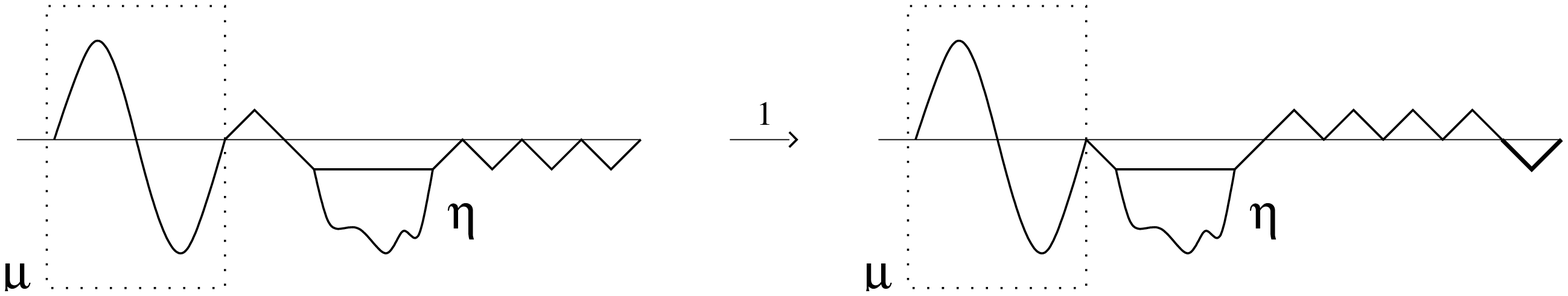, width =3.4in, clip=}
\caption{\small{The underground path generated by $\omega_{|0}$ in
the case 2.1)}} \label{fig:casob1}\vspace{-15pt}
\end{center}
\end{figure}

\item [2.2) $\eta  = \varepsilon$.] In this case, the underground
path obtained by means of the construction described in 2.1) is
$\omega' = \mu'\, \barx \,x\,(x\barx)^{j}\, \barx \, x$ and it
contains the forbidden pattern $\pat = (x\barx)^jx$ if $\mu'$ ends
with the sequence $(\barx x)^j$ or with the sequence $\barx \,
\eta' \, x\,(\barx x)^{j-1}$, where $\eta'$ is a not empty
strongly negative path. Let us take the longest suffix of
$\omega_{|0} =\mu'\, x\,\barx \,(\barx x)^{j}$ into account so
that $\omega_{|0} = \varphi \, \nu_1 \,  \nu_2 \ldots \nu_k$,
where
\begin{eqnarray*}
\nu_1&=&\barx \, \lambda \, x \, (\barx x)^{j-1} \, x \, \barx\\
\nu_i &=& (\barx x)^j \, x\, \barx \qquad \mbox{   $1 < i < k$}\\
\nu_k&=&(\barx x)^{j}
\end{eqnarray*}
and $\lambda$ is the empty path or is a strongly negative path.
Every sequence $\nu_i$, $1 \leq i \leq k$, will be changed into
$\bar{\nu_i}$ in the following way:
\begin{description}

\item[2.2.1)] if $\varphi$ is a path that does not end with a peak
$x\,\barx$, then
\begin{eqnarray*}
\bar{\nu}_1&=& (x\barx)^j\,\barx \, \lambda \,x\\
\bar{\nu}_i&=&(x \barx)^j\, \barx \, x \qquad \mbox{   $1 < i < k$}\\
\bar{\nu}_k&=&(x\barx)^j\,\barx \,x
\end{eqnarray*}
and the underground path generated by $\omega_{|0}$ is
$\varphi\,\bar{\nu}_1 \ldots \bar{\nu}_k$ (see
Figure~\ref{fig:casob2i});

\begin{figure}[!htb]
\begin{center}
\epsfig{file=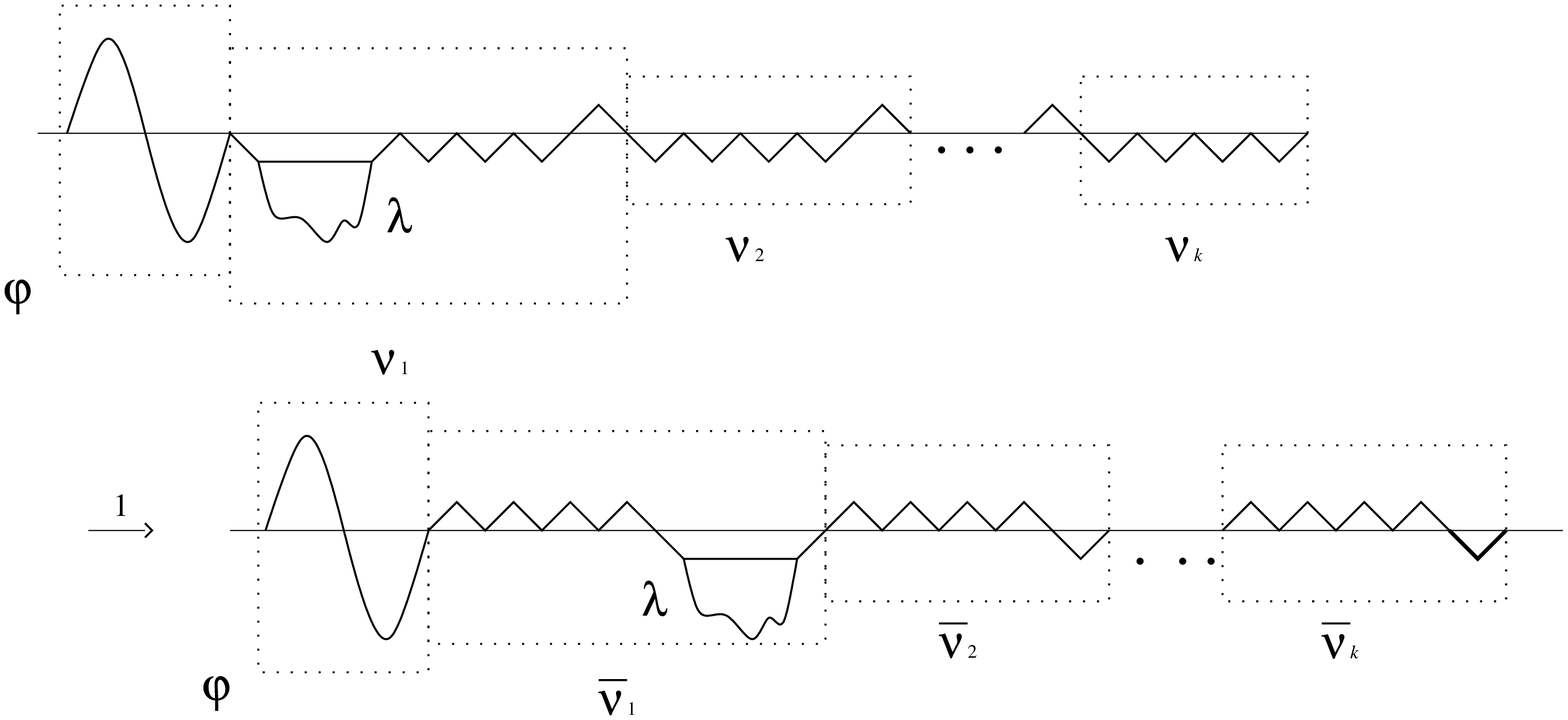, width =3.7in, clip=}
\caption{\small{The underground paths generated by $\omega_{|0}$
in the case 2.2.1)}} \label{fig:casob2i}\vspace{-15pt}
\end{center}
\end{figure}

\item[2.2.2)] if $\varphi$ ends with a peak $x\barx$, that is
$\varphi=\varphi'\,x\barx$, then
\begin{eqnarray*}
\bar{\nu}_1&=& \barx \, \lambda \,x(x\barx)^j\, \barx \,x  \\
\bar{\nu}_i&=&(x \barx)^j\, \barx \, x \qquad \mbox{   $1 < i < k$}\\
\bar{\nu}_k&=&(x\barx)^j\,\barx \,x
\end{eqnarray*}
and the underground path generated by $\omega_{|0}$ is
$\varphi'\,\bar{\nu}_1 \ldots \bar{\nu}_k$ (see
Figure~\ref{fig:casob2ii}).

\begin{figure}[!htb]
\begin{center}
\epsfig{file=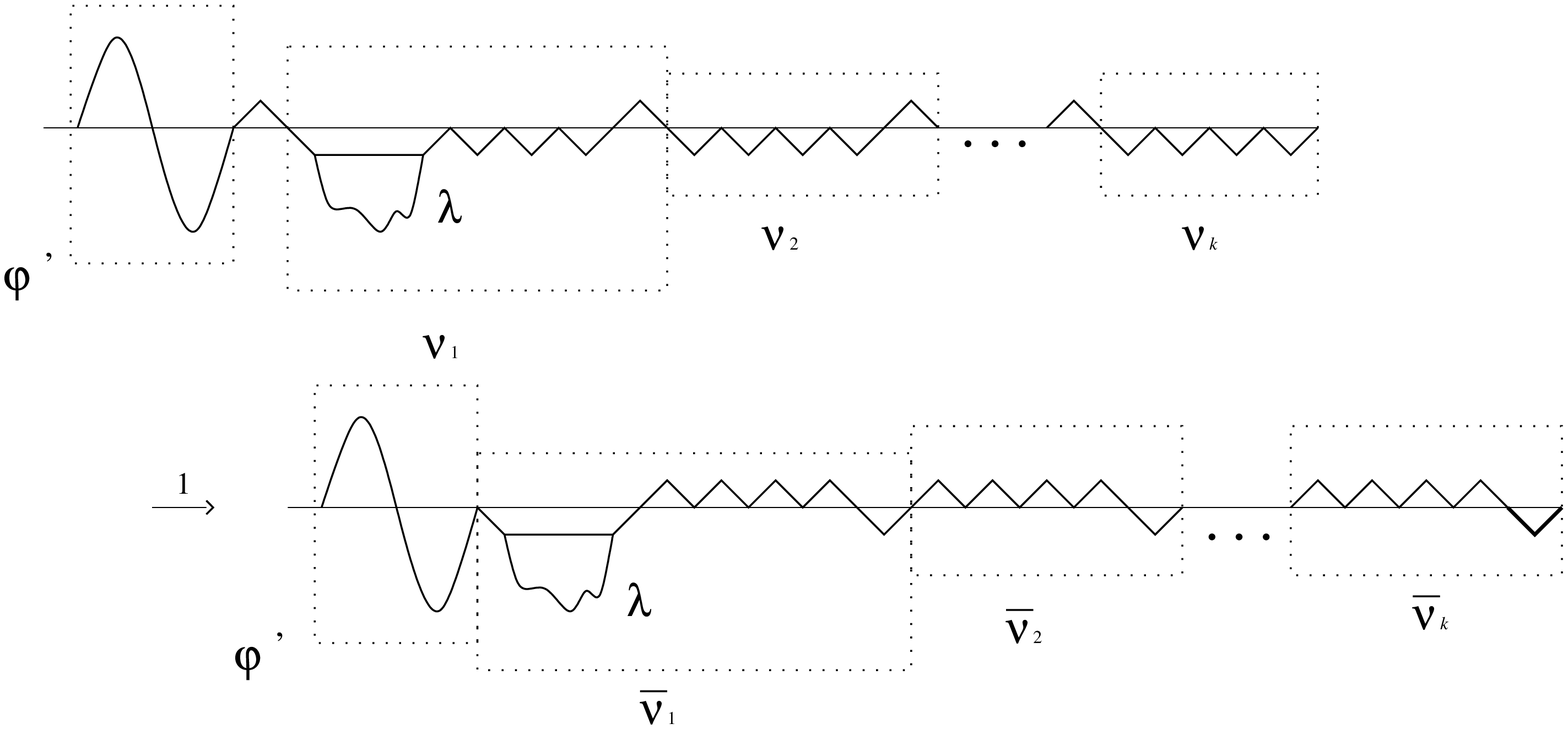, width =3.8in, clip=}
\caption{\small{The underground paths generated by $\omega_{|0}$
in the case 2.2.2)}} \label{fig:casob2ii}\vspace{-15pt}
\end{center}
\end{figure}

\end{description}
\end{description}

\end{description}

\subsection{The underground path generated by
$\omega_{|k}$}\label{sec:under} Now let us describe how to obtain
the underground path generated by $\omega_{|k}, k \geq 2$.

For any $h$, $1 \leq h \leq j$, let $\omega' =\nu\varphi$ be the
path obtained from  $\omega_{|k}$ and ending on the $x$-axis with
a positive suffix, $\varphi$ is the rightmost suffix in $\omega'$
which is primitive.

 If the path $\varphi^c$ does not contain the forbidden pattern $\pat$,
 the underground path generated by $\omega_{|k}$ is $\nu\varphi^c$.

 If the path $\varphi^c$ contains the forbidden pattern $\pat$, we must apply a \emph{swap}
 operation $\Phi$ in order to obtain a path $\varphi_1=\Phi(\varphi^c)$ avoiding
 the forbidden pattern. The underground path generated by $\omega_{|k}$ is $\nu\varphi_1$.

Before describing the $\Phi$ operation on $\varphi^c$, let us
consider the following proposition.
\begin{prop}\label{prop1}Let $\mu \in F^{[\pat]}$ a primitive path;
$\mu^c$ contains the forbidden pattern $\pat = (x\barx)^jx$ if and only
if $\mu$ contains the pattern $\mathfrak{p'}=(\barx)^2(x\barx)^j\barx$.
\end{prop}

From Proposition~\ref{prop1} it follows that, if $\varphi^c$
contains the forbidden pattern $\pat$, then it is preceded and
followed by at least a rise step.

\begin{figure}[!htb]
\begin{center}
\epsfig{file=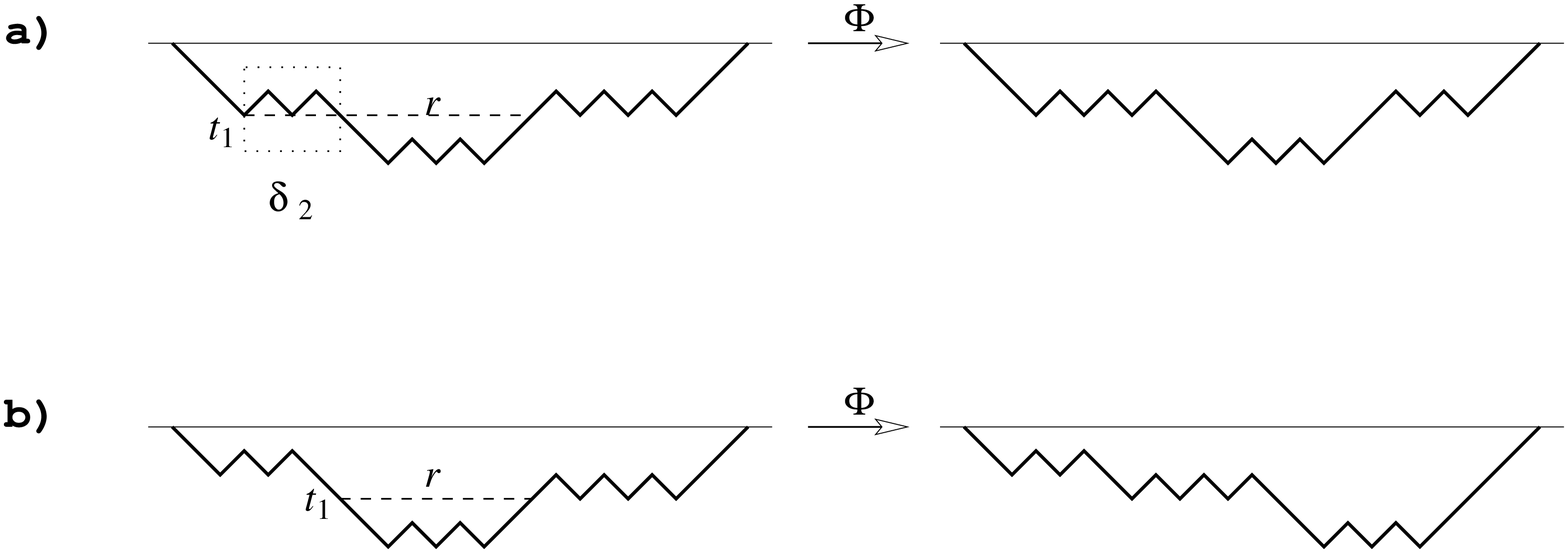, width =4in, clip=} \caption{\small{Some
examples of the $\Phi$ operation,
$\pat=(x\barx)^3x$}}
\label{fig:swap}
\end{center}
\end{figure}

Operation $\Phi$ must generate a path $\varphi_1$ avoiding the
forbidden pattern $\pat= (x\barx)^jx$ and such that $\varphi_1^c
\in F\backslash F^{[\pat]}$; in this way $\varphi_1$ is not the
complement of any path in $F^{[\pat]}$. The path $\varphi_1
=\Phi(\varphi^c)$ is obtained in the following way:
\begin{itemize}
\item [i)] consider the straight line $r$ from the beginning of
the
 pattern $\mathfrak{p} =(x\barx)^jx$ and let $t_1$ be the rightmost
 point in which $r$ intersects $\varphi^c$ on the left of $\pat$ such that $t_1$
 is preceded by at least two fall steps;

\item [ii)] let $\delta_2=(x\barx)^m$, $0 \leq m < j$, the
subsequence on the right of $t_1$,
 followed by at least a fall step;

\item [iii)] \emph{swap} the initial subsequence
$\delta_1=(x\barx)^j$ of $\pat$  and $\delta_2$. Let us remark
that $\delta_2$ can not be equal to $(x\barx)^j$ as $\varphi$ does
not contain the forbidden pattern $\pat =(x\barx)^jx$ (see
Figure~\ref{fig:swap}.a)). When $m=0$, that is $\delta_2$ is the
empty word, we simply insert $\delta_1$ into $t_1$ (see
Figure~\ref{fig:swap}.b)).
\end{itemize}
Operation $\Phi$ is applied to each forbidden pattern in
$\varphi^c$.

\begin{prop}\label{prop2} Let $\varphi_1 = \Phi(\varphi^c)$, then
$\varphi_1^c \in F\backslash F^{[\pat]}$.
\end{prop}
\noindent \textbf{Proof.} The $\Phi$ operation transforms the
subsequence $\varrho_1= (\barx)^m\,\delta_2\,\barx$, ($m \geq 2$),
of $\varphi^c$  into the subsequence $\varrho_2
=(\barx)^m\,\delta_1\,\barx =(\barx)^m\,(x\barx)^j\,\barx$ of
$\varphi_1$. The complement of $\varrho_2$ is
$$\varrho_2^c= (x)^{m}\,(\barx x)^j\,x\, = (x)^{m-1}\,(x\barx)^j\,x\,x$$
So $\varphi_1^c$ contains the forbidden pattern $\pat=(x\barx)^jx$.
\fine

\begin{prop} \label{prop3} Let $\mu \in F\backslash F^{[\pat]}$ a primitive
path such that $\mu^c \in F^{[\pat]}$. Then there exists a path
$\eta \in F^{[\pat]}$ such that $\mu^c =\Phi(\eta^c)$.
\end{prop}
\noindent \textbf{Proof.} If $\mu \in F\backslash F^{[\pat]}$ and
$\mu^c \in F^{[\pat]}$ then $\mu^c$ contains the pattern $\barx
\barx(x\barx)^j\barx$; we apply to $\mu^c$ the following operation
$\Phi^{-1}$:
\begin{itemize}
\item [i)]consider the straight line $r$ from the end of the
 pattern $(x\barx)^j$ and let $t_2$ be the leftmost
 point where $r$ intersects $\mu^c$ on the right of $(x\barx)^j$ such that $t_2$
 is followed by at least two rise steps;

\item [ii)]let $\delta_2=(x\barx)^m$, $0 \leq m < j$, the
subsequence on the left of $t_2$, preceded by at least a rise
step;

\item [iii)] swap the subsequence $(x\barx)^j$ and $\delta_2$.
When $m=0$, that is $\delta_2$ is the empty word, we simply insert
$(x\barx)^j$ into $t_2$.
\end{itemize}
\fine

\begin{figure}[!htb]
\begin{center}
\epsfig{file=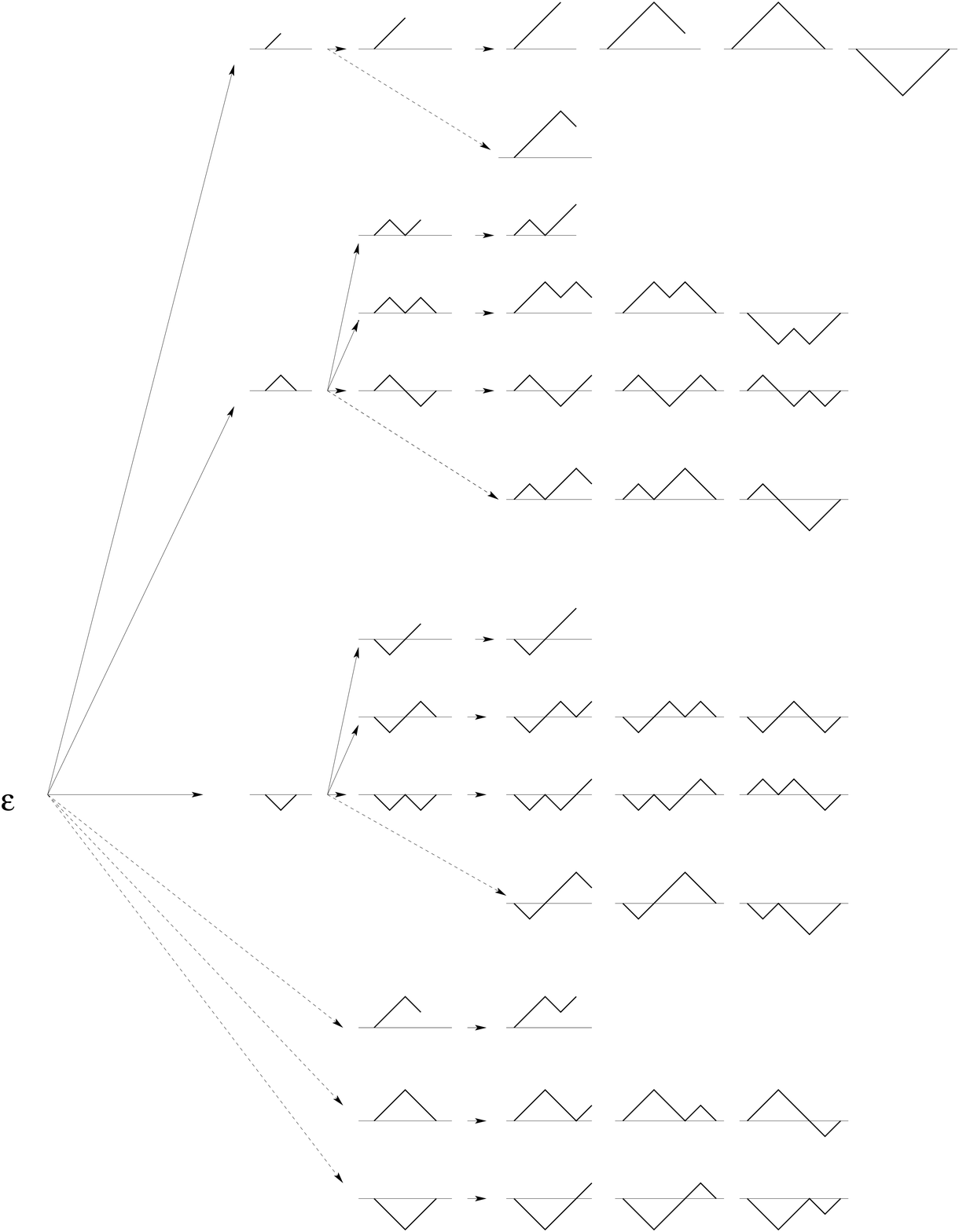, width =4in, clip=}\caption{\small{The
initial steps of the generating algorithm of the paths
corresponding to words in $F^{[\pat]}$, $\pat = (x\barx)^2x =
(10)^21$. Dotted lines are related to $h=2$} }\label{fig:tree}
\end{center}
\end{figure}

\noindent Figure~\ref{fig:tree} shows the initial steps of the
generating algorithm of the paths corresponding to words in
$F^{[\pat]}$, $\pat = (x\barx)^2x = (10)^21$.

Let us remark that, following the above constructions, given a
path $\omega$, the number of generated paths depends only on the
ordinate of endpoint of $\omega$.

So, the complete generating algorithm can be briefly described by
the succession rule (\ref{one}) (for more details on succession
rules see \cite{BDPP99,CORTE00,FPPR03})
\begin{equation}\label{one}
 \left\{
\begin{array}{ll}
   (0)\\
   (0)\stackrel{h}{\rightsquigarrow}(0)(0)(1)& 1 \leq h \leq j\\
   (1)\stackrel{h}{\rightsquigarrow}(2)& 1 \leq h \leq j\\
   (k)\stackrel{h}{\rightsquigarrow}(0)(0)(1)\cdots(k-1)(k+1)& 1 \leq h \leq j, \, k \geq 2\\

\end{array}
\right.
\end{equation}

\noindent where each number corresponds to the ordinate of the
endpoint of a path. The zero in the first line in (\ref{one}) is
associated with the empty path. The second line in (\ref{one}) is
associated with operations $(1)$ and $(4)$, the third line is
associated with operations $(2)$ and $(5)$, and the last line
describes the construction when the endpoint has ordinate $k \geq
2$, underground path included.

\section{Exhaustive generation}\label{sec:proof} In this section we prove that
the construction described in Section 3 allows to generate the
class $F^{[\pat]}$ exhaustively for any fixed forbidden pattern
$\pat = (10)^j1$, $j \geq 1$, in the sense that all the words in
$F^{[\pat]}$ with $n$ 1's, $n \geq 0$, can be generated.

\begin{teo} \label{teo:uno} Given a fixed forbidden pattern $\pat = (10)^j1$, $j \geq
1$, the construction described in Section~\ref{sec:algo} generates
all the paths with $n$, $n \geq 0$, rise steps representing the
binary words in $F^{[\pat]}$ with $n$ 1's.
\end{teo}

 Let $\omega \in F^{[\pat]}$, then
\begin{equation}
\omega= \varphi_0\varphi_1\varphi_2 \ldots \varphi_s
\end{equation}
that is, $\omega$ is made of $s+1$ sub-paths such that:
\begin{itemize}
\item $\varphi_0$ is the empty path $\varepsilon$,

\item  $\varphi_i$, $1 \leq i <s$ is a path in $F^{[\pat]}$
beginning from and ending on the $x$-axis,

\item $\varphi_s$ is a path in $F^{[\pat]}$ beginning from the
$x$-axis with endpoint at ordinate $k \geq 0$.
\end{itemize}
The proof of Theorem~\ref{teo:uno} is obtained by induction on the
number of sub-paths.

\noindent \textbf{Proof.} The empty path $\varepsilon$ is
generated as the starting point of the algorithm. Let us assume
that all the possible sub-paths $\varphi_1\varphi_2 \ldots
\varphi_i$ of $\omega$ are generated. We prove that the algorithm
generates the sub-path $\varphi_1\varphi_2 \ldots
\varphi_i\varphi_{i+1}$ for any path $\varphi_{i+1} \in
F^{[\pat]}$.

Let $\varphi_i$ be a path that does not end with the pattern
$(x\barx)^j$. In this case the path $\varphi_{i+1}$ may be either
positive or negative.

If $\varphi_{i+1}$ is a positive path we have to prove that all
the positive paths are generated and this will be demonstrated in
Section~\ref{sec:positive}. In the case of a negative path,
denoting  $d$ the largest value of its absolute ordinate, let us
remark that:

\begin{itemize}
\item the negative paths with $d=1$ are $(\barx x)^\ell$, $1 \leq
\ell \leq j$, and they are generated by iterating the construction
(\ref{alfa});

\item negative paths with $d=2$ are $\barx (\barx x)^\ell x$, $1
\leq \ell \leq j$, and they are generated by means of (\ref{alfa})
when $1 \leq \ell < j$, or by means of (\ref{alfa1}) when
$j=\ell$;

\item let  $\gamma$ be a negative path with $d>2$; if $\gamma^c
\in F^{[\pat]}$ then $\gamma$ is the underground path generated by
a positive path with endpoint at ordinate $k \geq 2$, (see
Section~\ref{sec:under}), otherwise $\gamma = \Phi(\eta^c)$ for a
positive path $\eta$ in $F^{[\pat]}$ with endpoint at ordinate $k
\geq 2$ (see Proposition~\ref{prop3}).
\end{itemize}

Note that when $j=1$ and $\varphi_i$ ends with the pattern $\barx
x$, the only possible negative path $\varphi_{i+1}$ with $d=2$ is
$\barx\barx xx$ and it is generated by applying the construction
(\ref{alfa1}) to the path $\varphi_ix\barx$.

When the suffix of $\varphi_i$ is the pattern $(x\barx)^j$,
$\varphi_{i+1}$ must be a negative path and the path
$\varphi_1\varphi_2 \ldots \varphi_i\varphi_{i+1}$ is the
underground path obtained by means of the construction described
in case 1) in Section~\ref{sec:negativesuffix} (see
Figure~\ref{fig:casoa}).

In the same way, when the sub-path $\varphi_i \varphi_{i+1}$ is of
type
$$(x\barx)^j\barx \lambda x((x\barx)^j\barx x)^r, \qquad r \geq 1 $$ or $$\barx \lambda x((x\barx)^j\barx
x)^r, \qquad r \geq 1 $$ where $\lambda$ is the empty path
$\varepsilon$ or is a strongly negative path, then it is generated
by the constructions described in case 2) in
Section~\ref{sec:negativesuffix} (see Figures~\ref{fig:casob1},
\ref{fig:casob2i} and \ref{fig:casob2ii}).

Then, if we show that all the possible positive paths are
generated, then we can claim that Theorem~\ref{teo:uno} is proved.

Moreover, we observe that for each path $\omega$ in $F^{[\pat]}$
with $n$ rise steps there exists one and only one path $\omega'$
in $F^{[\pat]}$ with $n-h$ rise steps, $1 \leq h \leq j$, such
that $\omega$ is obtained from $\omega'$ by means of the
construction described in Section~\ref{sec:algo}.

This assertion is a direct consequence of the construction, since
the actions described are univocally determined. \fine

\subsection{Positive paths}\label{sec:positive}
In this section we prove that all the positive paths with $n$ rise
steps are generated by means of the construction described in
Section~\ref{sec:algo}. In the sequel of this section we analyze
only \emph{positive} paths.

The proof is obtained by induction on $n$. There are only two
paths with $n=1$ rise step, that is $x$ and $x\barx$, and they are
generated by means of construction (\ref{alfa}) applied to the
empty path $\varepsilon$. Let us assume that all the paths with
$n' <n$ rise steps are generated; we will prove that all the paths
with $n$ rise steps are generated.

Note that, following the construction given in \cite{BDPP99}, a
path with $n$ rise steps can be obtained from a Dyck path $\omega$
with $n-1$ rise step by inserting one rise step in each point at
ordinate $i$ of its last descent followed by $q$ fall steps, $0
\leq q \leq i+1$. We will prove that all the paths obtained so are
also generated following the constructions given in the above
sections. Let us denote by $\omega_q$ a paths ending with $q$ fall
steps.

Let $m$ be the number of fall steps in the last descent of
$\omega$. First of all, we note that for any value of $m$ the
paths obtained by inserting a rise step in the point at ordinate 0
are generated by means of the constructions (\ref{alfa}) or
(\ref{alfa1}) applied to the path $\omega$. Here we give the proof
for the case with $m >2$, distinguishing three cases: $i=1$, $(1<
i < m-1)\vee (i=m)$ and $i=m-1$. The analogous and simple cases
$m=1$ and $m=2$ are left to the reader.


\begin{itemize}
\item $i =1$. Let $\omega = \gamma_{|m-1}x(\barx)^m$. The
insertion of a rise step in the point at ordinate 1 gives three
paths:
\begin{itemize}
\item $\omega_0=\gamma_{|m-1}x(\barx)^{m-1}x$, which is generated
by means of (\ref{beta}) applied to the prefix
$\gamma_{|m-1}x(\barx)^{m-1}$ of $\omega$,

\item $\omega_1 =\gamma_{|m-1}x(\barx)^{m-1}x\barx$ and $\omega_2
=\gamma_{|m-1}x(\barx)^{m-1}x\barx\barx$, which are the paths with
endpoints at ordinate 1 and 0, respectively. If $j >1$, then
$\omega_1$ and $\omega_2$ are generated by the construction
(\ref{gamma}), where $k=m-1$, applied to the path $\gamma_{|m-1}$
with $h=2$, otherwise, if $j=1$, they are generated by means of
(\ref{gamma1}) with $k=m-1$ and $h=1$ applied to the path
$\gamma_{|m-1}x\barx$.
\end{itemize}
\item $(1< i < m-1)\vee (i=m)$. The insertion of a rise step in
the point at ordinate $i$ gives $i+2$ paths $\omega_q$, $0 \leq q
\leq i+1$. The paths $\omega_q$ with $q\neq 1$ are all the
positive paths generated by means of (\ref{gamma}) with $k=i$ and
$h=1$ applied to the prefix of $\omega$  of length $|\omega|-i$.
The path $\omega_1 =\gamma_{|m-1}x(\barx)^{m-i}x\barx$ is the path
with endpoint at ordinate $i$. When $ j >1$, $\omega_1$ is
generated by means of construction (\ref{gamma}) with $k=m-1$ and
$h=2$ applied to the path $\gamma_{|m-1}$, while, when $j=1$, it
is generated by means of (\ref{gamma1}), with $k=m-1$ and $h=1$,
applied to the path $\gamma_{|m-1}x\barx$ .

\item $i=m-1$. The insertion of a rise step in the point at
ordinate $m-1$ generates $m+1$ paths $\omega_q$, $0 \leq q \leq
m$. The paths $\omega_q$ with $q\neq 1$ are all the positive paths
generated by means of (\ref{gamma}) with $k=m-1$ and $h=1$ applied
to the prefix of $\omega$  of length $|\omega|-m+1$. As far as the
generation of $\omega_1$ is concerned, we have to distinguish
three cases:
\begin{enumerate}
\item If $\omega=\gamma_{|m-2}x(x\barx)^{\ell-1}(\barx)^{m-1}$, $1
\leq \ell-1<j$, then $\omega_1 =\gamma_{|m-2}x(x\barx)^{\ell}$. If
$\ell < j$, then $\omega_1$ is the path with endpoint at ordinate
$(k+1)$ generated by means of (\ref{gamma}) with $k=m-2$ and $
h=\ell+1$, applied to the path $\gamma_{|m-2}$. If $\ell=j$, then
$\omega_1$ is the path with endpoint at ordinate $(k+1)$ generated
by means of (\ref{gamma1}) with $k=m-2$ and $h=1$ applied to the
path $\gamma_{|m-2}(x\barx)^{\ell}$. Note that, when $m=3$, the
endpoint of the prefix $\gamma_{|m-2}$ has ordinate 1, and the
path $\omega_1$ is obtained applying the construction (\ref{beta})
(or (\ref{beta1})) instead of (\ref{gamma}) (or (\ref{gamma1})).

\item If
$\omega=\gamma_{|m+m'-2}x(\barx)^{m'}(x\barx)^{\ell-1}(\barx)^{m-1}$,
$1 \leq \ell-1<j$ and $m' > 2$, then $\omega_1
=\gamma_{|m+m'-2}x(\barx)^{m'}(x\barx)^{\ell}$. If $\ell < j$,
then $\omega_1$ is the path with endpoint at ordinate $(k-m'+1)$
generated by means of (\ref{gamma}) with $k=m+m'-2$ and $h=\ell+1$
applied to the path $\gamma_{|m+m'-2}$. If $\ell=j$, then
$\omega_1$ is generated by means of (\ref{gamma1}) with $k=m+m'-2$
and $h=1$ applied to the path $\gamma_{|m+m'-2}(x\barx)^{\ell}$.

\item If
$\omega=\gamma_{|m-2}(x\barx)^rx(x\barx)^{\ell-1}(\barx)^{m-1}$,
$0 \leq r <j$ and $1 \leq \ell-1<j$, then $\omega_1
=\gamma_{|m-2}(x\barx)^rx(x\barx)^{\ell}$. If $\ell < j$, then
$\omega_1$ is the path with endpoint at ordinate $(k+1)$ generated
by means of (\ref{gamma}), with $k=m-2$ and $h=\ell+1$ applied to
the path $\gamma_{|m-2}(x\barx)^r$. If $\ell=j$, then $\omega_1$
is generated by means of (\ref{gamma1}), with $k=m-2$ and $h=r+1$
applied to the path $\gamma_{|m-2}(x\barx)^\ell$.

\end{enumerate}

\end{itemize}

\fine

\section{Conclusions and further developments}
In this paper we propose an algorithm for the construction of
particular binary words, according to the number of 1's, excluding
a fixed pattern $\mathfrak{p}=(10)^j1$, $j \geq 1$.

Successive studies should take into consideration binary words
avoiding different forbidden patterns both from an enumerative and
a constructive point of view.

Moreover, it would be interesting to study words avoiding patterns
which have a different shape, that is not only patterns consisting
of a sequence of rise and fall steps. This could be the first step
in the study of a possible universal generating algorithm for
pattern avoiding words.

Another interesting field of study is to determine a sort of
invariant class of avoiding patterns that is the paths $\pat_1,
\pat_2, \dots , \pat_l$ such that $|F^{[\pat_1]}|=|F^{[\pat_2]}|=
\dots =|F^{[\pat_l]}|$ with consequent bijective problems.

One could also consider a forbidden pattern on an arbitrary
alphabet and investigate words avoiding that pattern, or study
words avoiding more than one pattern and the related combinatorial
objects, considering various parameters.


\begin{thebibliography}{99}

\bibitem{AA02} A. Apostolico, M. Atallah, Compact recognizers of
episode sequences, \emph{Information and Computation} 174(2)
(2002) 180--192.

\bibitem{BDPP99} E. Barcucci, A. Del Lungo, E. Pergola, R.
Pinzani,  ECO: a methodology for the enumeration of combinatorial
objects,  \emph{J. Difference Equ. Appl}. 5 (1999) 435--490.

\bibitem{BMPP11} S. Bilotta, D. Merlini, E. Pergola, R. Pinzani,
Binary words avoiding a pattern and marked succession rule,
\emph{Lattice Path Combinatorics and Applicatons, Siena, July 4-7,
2010} (available on line arXiv:1103.5689), (2010).


\bibitem{CORTE00} S. Corteel, S\'{e}ries g\'{e}n\'{e}ratrices exponentielles
pour les eco-syst\`{e}mes sign\'{e}s, \emph{Proc. of the 12th
International Conference of Formal Power Series and Algebraic
Combinatorics, Moscow}, (2000).

\bibitem{FPPR03} L. Ferrari, E. Pergola, R. Pinzani, S.
Rinaldi, Jumping succession rules and their generating functions,
\emph{Discrete Math.} 271 (2003) 29--50.

\bibitem{FSV06} P. Flajolet, W. Szpankowski, B. Valle, Hidden word
statistics, \emph{Journal of the ACM} 53(1) (2006) 147--183.

\bibitem{GO80} L.J. Guibas, M. Odlyzko, Long repetitive patterns
in random sequences, \emph{Zeitschrift f\"{u}r
Wahrscheinlichkeitstheorie} 53 (1980) 241--262.

\bibitem{GO81} L.J. Guibas, M. Odlyzko, String overlaps, pattern
matching, and nontransitive games, \emph{Journal of Combinatorial
Theory, Series A} 30 (1981) 183--208.

\bibitem{KS94} S. Kumar, E.H. Spafford, A pattern matching model
for misuse intrusion detection, In \emph{Computer Security},
(1994) 11--21.


\bibitem{SF95} R. Sedgewick, P. Flajolet, An Introduction to the
Analysis of Algorithms, Chapman-Hall, London (1995).



\end{thebibliography}
\end{document}